\documentclass[letter,oneside,9pt,twocolumn,article]{aiaa}
\usepackage{rotating}
\usepackage{bm}
\usepackage{flushend}
\usepackage[top=0.5in, bottom=1.0in, left=0.7in, right=0.7in]{geometry}
\usepackage[bf,footnotesize]{subfigure}
\usepackage{adjustbox}
\usepackage{tabularx}
\usepackage{multirow}
\usepackage{multicol}
\usepackage{indentfirst} 
\usepackage[hidelinks]{hyperref}
\usepackage{float}
\usepackage{xcolor}
\usepackage[super,comma,sort,compress]{natbib}

\usepackage{mwe}
\usepackage[markcase=noupper]{scrlayer-scrpage}
\usepackage{caption} 

\ihead*{{\it Journal of Chemical Physics 164, 181101 (2026), \href{https://doi.org/10.1063/5.0325075}{10.1063/5.0325075}}}
\ohead*{\pagemark}
\ifoot*{}% left side of footer
\ofoot*{\rm Unrestricted content}% right side of footer
\newcommand{\alb}{\vspace{0.1cm}\\} % array line break

\author{
Bernard Parent\thanks{Associate Professor, bparent@arizona.edu.} 
\\[0.3em] \it University of Arizona, Tucson, AZ 85721, USA.
}

\title{
Superelastic Heating in Treanor-Gordiets Plasmas: A Unified Analytic Closure
}

\abstract{ 
In thermally non-equilibrium plasmas, conventional harmonic models
can significantly mispredict  superelastic electron heating rates. When the
vibrational temperature exceeds the gas temperature ($T_{\rm v}>T_{\rm g}$), these models
underestimate energy transfer by several times; conversely, they
overestimate heating when $T_{\rm g}>T_{\rm v}$. We show that this discrepancy
arises from neglecting the exponential heating from overpopulated,
high-lying states in anharmonic Treanor-Gordiets distributions, and
their thermodynamic depopulation at high gas temperatures. To resolve
this, we derive a closed-form, thermodynamically consistent
macroscopic closure based on detailed balance and a second-order Dunham
expansion. This unified framework introduces an analytic anharmonic
correction factor that captures the kinetic competition between
vibrational-vibrational (V-V) up-pumping and vibrational-translational (V-T)
relaxation. By predicting the Treanor
minimum, this formulation recovers the fidelity of full state-to-state
kinetic benchmarks. Ultimately, this model provides a governing equation for heat exchange between electrons and excited states in non-equilibrium environments---including plasma-assisted combustion and hypersonic flows---enabling the development of accurate, rate-limited reduced-order models for macroscopic fluid solvers.\\
~\\
}

\nomenclature{
\begin{nomenclaturelist}{Roman symbols}
\item[$C_k$]{charge of $k$th  species, C}
  \end{nomenclaturelist}
  \begin{nomenclaturelist}{Greek symbols}
\item[$\beta_{k}^{\rm n}$]{parameter equal to 1 when $k$th species is neutral, 0 otherwise}
  \end{nomenclaturelist}
}

\begin{document}
\maketitle
\dropword
Electron-vibrational (e-V) energy coupling is a cornerstone process in a diverse array of applications, including hypersonic flight, plasma-assisted combustion (PAC), laser-induced plasmas (LIP), and emerging technologies in green energy and plasma medicine. In post-shock re-entry, it affects plasma density  on which various technologies such as electromagnetic shielding,\cite{dtic:1992:gregorie} electron transpiration cooling,\cite{aip:2017:hanquist,aiaapaper:2021:parent} and magnetohydrodynamic flow control\cite{aiaaconf:2022:moses,jtht:2023:parent,aiaa:2025:parent} depend. In LIP\cite{jap:2019:peters,jap:2010:shneider,jap:2023:pokharel,book:1989:radziemski,jpd:2019:alberti,jcp:2020:munafo} and in PAC,\cite{apl:2011:bak,jtpht:1998:adamovich,rsta:2015:adamovich,jap:2023:chen,cf:2016:castela,cf:2025:dijoud} superelastic heating dominates the inter-pulse phase, while in semiconductor and biomedical discharges,\cite{jap:2000:kushner,jpd:2007:arakoni,psst:2022:jiang,jap:2024:miyake,psst:2016:bruggeman,psst:2023:davies} it controls process optimization and effluent reactivity.

Fundamentally, electron-vibrational coupling dictates the electron temperature, $T_{\rm e}$. Within macroscopic plasma fluid models, $T_{\rm e}$ evolves according to the electron energy transport equation, where the net energy transfer is driven by two competing mechanisms: a macroscopic cooling rate via vibrational excitation ($Q_{\rm e-v}$) acting as a heat sink, and a reverse superelastic heating rate ($Q_{\rm v-e}$) acting as a heat source. It is this mutual heat exchange between excited states and energetic electrons that ultimately dictates how these two originally independent bodies couple into alignment. However, a critical asymmetry exists in our current modeling capabilities: while $Q_{\rm e-v}$ is well-characterized by established cross-sections, the superelastic heating term $Q_{\rm v-e}$ remains poorly understood.

Despite its importance, accurately modeling the vibrational-electron heating rate has presented significant modeling challenges. Early phenomenological models typically scaled the heating rate using simple temperature or energy ratios.\cite{jap:2010:shneider,jtht:2012:kim,jtht:2013:farbar} While these formulations successfully drive the system toward equilibrium (by forcing $T_{\rm e}$ to equal the vibrational temperature $T_{\rm v}$), they lack a rigorous kinetic derivation. Conversely, more recent models founded on the principle of detailed balance\cite{jap:2019:peters,jap:2023:pokharel} do not always ensure convergence of the electron and vibrational temperatures at thermal equilibrium, which leads to a violation of thermodynamic consistency.\cite{pf:2025:rodriguez:2}

To resolve these discrepancies, we recently introduced a thermodynamically consistent model for vibrational-electron heating.\cite{pf:2025:rodriguez:2} By enforcing strict adherence to the second law of thermodynamics, this model guaranteed the correct equilibration of $T_{\rm e}$ and $T_{\rm v}$. However, its derivation was restricted to mono-quantum transitions and assumed a Boltzmann distribution, limiting its validity to low-temperature regimes ($T_{\rm e} \lesssim 1.5$ eV). We subsequently addressed the energy limitation in a second study,\cite{pf:2026:parent} which generalized the framework to include multi-quantum overtone transitions. This extension allowed the model to be applied to high-energy discharges typical of PAC and LIP. 

Nonetheless, a critical limitation remains: both prior formulations rely on the assumption of a Boltzmann distribution of vibrational states. This assumption breaks down in highly non-equilibrium regimes driven by rapid thermal energy increase---whether through strong shock heating in hypersonic flows\cite{pnas:2018:singh,jcp:2014:parsons} or intense electron impact in plasma discharges. This deviation is particularly severe in conditions where the vibrational temperature exceeds the gas temperature ($T_{\rm v} > T_{\rm g}$), such as in the recombination wake of hypersonic vehicles, the inter-pulse stages of PAC, or in the non-equilibrium discharges used for chemical synthesis. In these scenarios, the electron energy is preferentially pumped into the vibrational modes while the translational mode remains cold, leading to a non-equilibrium Treanor distribution characterized by the overpopulation of high-energy states.\cite{jcp:1968:treanor} Neglecting this anharmonic distortion can lead to substantial errors in the predicted heating rates.

In this Communication, we derive a unified thermodynamic framework for electron-vibrational energy coupling. By enforcing detailed balance upon the second-order Dunham expansion, we establish an analytical governing equation for the anharmonic Treanor-Gordiets regime. This formulation resolves the limitations of conventional harmonic models, which underestimate heating, while incorporating the kinetic cutoff necessary to prevent the unphysical divergence of pure Treanor scalings. The result is a thermodynamically consistent formulation of energy transfer applicable to diatomic systems driven far from equilibrium.

The Dunham expansion for the energy of a vibrational level $n$ is typically described by the spectroscopic series:\cite{pr:1932:dunham,book:1950:herzberg}
\begin{align}
    {\mathcal E}_n = hc~ \bigg[ &~\omega_{\rm e}(n + 0.5) - \omega_{\rm e}x_{\rm e}(n + 0.5)^2 \nonumber\\ &+ \omega_{\rm e}y_{\rm e}(n + 0.5)^3 + \omega_{\rm e}z_{\rm e}(n + 0.5)^4 + \dots \bigg]
    \label{eq:dunham_full}
\end{align}
where $h$ is the Planck constant, $c$ is the speed of light, and $\omega_{\rm e}$, $\omega_{\rm e}x_{\rm e}$, $\omega_{\rm e}y_{\rm e}$, and $\omega_{\rm e}z_{\rm e}$ are the harmonic frequency and the consecutive anharmonic corrections. The energy gap $\Delta {\mathcal E}_{n,m}$ for a transition from level $n$ to $n+m$ is the difference ${\mathcal E}_{n+m} - {\mathcal E}_n$ where $m$ is the change in vibrational quantum number. 

If we truncate the series after the second anharmonic correction (neglecting terms of order $\omega_{\rm e}z_{\rm e}$ and higher), the energy gap is given by the  algebraic difference:
\begin{align}
    \Delta {\mathcal E}_{n,m} = hc~ \bigg[ &~m\omega_{\rm e} - m\omega_{\rm e}x_{\rm e}(2n + m + 1)\nonumber\\ 
    &+ \omega_{\rm e}y_{\rm e} \left( (n+m+0.5)^3 - (n+0.5)^3 \right) \bigg]
    \label{eq:standard_gap}
\end{align}
In fluid models, the vibrational energy scale is anchored by the fundamental characteristic vibrational temperature $\theta_{\rm v} \equiv \Delta {\mathcal E}_{0,1}/k_{\rm B}$ with $k_{\rm B}$ the Boltzmann constant. To link the theoretical harmonic frequency $\omega_{\rm e}$ to this macroscopic parameter, we evaluate Eq.~(\ref{eq:standard_gap}) for the fundamental transition ($n=0$, $m=1$). Noting that the cubic term difference is $(1.5)^3 - (0.5)^3 = 3.25$, we obtain:
\begin{equation}
    k_{\rm B}\theta_{\rm v} = hc\omega_{\rm e}(1 - 2x_{\rm e} + 3.25 y_{\rm e})
\end{equation}
with $x_{\rm e}$ and $y_{\rm e}$ the first and second anharmonic corrections. Inverting this relation to express $\omega_{\rm e}$ in terms of $\theta_{\rm v}$ leads to:
\begin{equation}
    hc\omega_{\rm e} = \frac{k_{\rm B}\theta_{\rm v}}{1 - 2x_{\rm e} + 3.25 y_{\rm e}}
    \label{eq:we_inversion}
\end{equation}
We substitute Eq.~(\ref{eq:we_inversion}) back into the general gap expression Eq.~(\ref{eq:standard_gap}). By algebraically factoring the numerator to  extract the denominator $(1 - 2x_{\rm e} + 3.25 y_{\rm e})$ we obtain:
\begin{align}
    \Delta {\mathcal E}_{n,m} = m k_{\rm B}\theta_{\rm v} \Bigg[ & 1 - \frac{x_{\rm e}(2n + m - 1)}{1 - 2x_{\rm e} + 3.25 y_{\rm e}} \nonumber \\
    & + \left. \frac{y_{\rm e} \left( (n+m+0.5)^3 - (n+0.5)^3 - 3.25m \right)}{m(1 - 2x_{\rm e} + 3.25 y_{\rm e})}  \right]
\end{align}
This yields the following form of the energy gap used in our model:
\begin{equation}
    \Delta {\mathcal E}_{n,m} = m k_{\rm B} \theta_{\rm v} \left[ 1 - \delta(n,m) \right]
    \label{eq:energy_gap_anh}
\end{equation}
where the anharmonic defect function $\delta(n,m)$ algebraically captures the deviation from harmonicity up to the second order, including both the first ($x_{\rm e}$) and second ($y_{\rm e}$) anharmonic corrections:
\begin{equation}
    \delta(n,m) = \frac{x_{\rm e}(2n + m - 1) - y_{\rm e} \left( 3n^2 + 3nm + m^2 + 3n + 1.5m - 2.5 \right)}{1 - 2x_{\rm e} + 3.25 y_{\rm e}}
    \label{eq:delta_nm_def}
\end{equation}
With the anharmonic energy gap established, we now integrate it into the kinetic formulation. As in the harmonic derivation,\cite{pf:2026:parent} the total cooling rate $Q_{\rm e-v}$ is defined as the sum over all transition channels $m$, where $m$ represents the change in vibrational quantum number:
\begin{equation}
    Q_{\rm e-v} = \sum_{m=1}^\infty Q_{\rm e-v}^{(m)}
    \label{eq:total_cooling_anh}
\end{equation}
where $Q_{\rm e-v}^{(m)}$ represents the macroscopic electron energy loss due to the specific transition channel $m$. 
The macroscopic cooling rate for channel $m$ is obtained by summing the contributions from all vibrational levels $n$:
\begin{equation}
    Q_{\rm e-v}^{(m)} = N_{\rm e} \sum_{n=0}^{\infty} N_n k_{n \to n+m} \Delta {\mathcal E}_{n,m}
    \label{eq:cooling_def_anh}
\end{equation}
where $N_{\rm e}$ is the electron number density, $N_n$ is the population density of vibrational level $n$, and $k_{n \to n+m}$ is the rate coefficient for electron-impact excitation from level $n$ to $n+m$. For the subsequent derivation, it is necessary to define the level-specific cooling flux $Q_{\rm e-v}^{(n,m)}$, which represents the cooling contribution from the specific transition $n \to n+m$:
\begin{equation}
    Q_{\rm e-v}^{(n,m)} \equiv N_{\rm e} N_n k_{n \to n+m} \Delta {\mathcal E}_{n,m}
    \label{eq:level_cooling_def}
\end{equation}
The corresponding macroscopic heating rate $Q_{\rm v-e}^{(n,m)}$ for the reverse superelastic processes (de-excitation from $n+m \to n$) is given by:
\begin{equation}
    Q_{\rm v-e}^{(n,m)} = N_{\rm e}  N_{n+m} k_{n+m \to n} \Delta {\mathcal E}_{n,m}
    \label{eq:heating_def_anh}
\end{equation}
where $N_{n+m}$ is the population of the upper state and $k_{n+m \to n}$ is the de-excitation rate coefficient. To relate these forward and reverse rates, we apply the principle of detailed balance. The rate coefficient for the reverse process is related to the forward process by the exponential of the specific anharmonic energy gap at the electron temperature $T_{\rm e}$:
\begin{equation}
    k_{n+m \to n} = k_{n \to n+m} \exp\left(\frac{\Delta {\mathcal E}_{n,m}}{k_{\rm B} T_{\rm e}}\right)
    \label{eq:detailed_balance_anh}
\end{equation}
The principle of detailed balance is applied in Eq.~(\ref{eq:detailed_balance_anh}) to determine the reverse rate coefficients. This step is required to ensure the thermodynamic consistency of the formulation; utilizing independent forward and reverse rates would fail to satisfy the second law of thermodynamics at equilibrium. While this relation formally implies a Maxwellian Electron Energy Distribution Function (EEDF), the final anharmonic correction factor derived here is expected to remain a very good estimate even when the electron distribution is non-Maxwellian. The physical justification for this extended applicability will be discussed further below Eq.~(\ref{eq:Qve_final}).

Furthermore, we assume a Treanor distribution for the vibrationally excited states. This distribution accounts for the overpopulation of high-energy states when the vibrational temperature $T_{\rm v}$ exceeds the translational gas temperature $T_{\rm g}$. For a Treanor distribution, the population of a specific vibrational level $n$ becomes:\cite{jcp:1968:treanor}
\begin{equation}
    N_n = N_0 \exp\left( - \frac{n {\mathcal E}_1}{k_{\rm B} T_{\rm v}} + \frac{n {\mathcal E}_1 - {\mathcal E}_n}{k_{\rm B} T_{\rm g}} \right)
    \label{eq:treanor_std}
\end{equation}
where $N_0$ is the ground state population, ${\mathcal E}_1$ is the energy of the first vibrational level, and ${\mathcal E}_n$ is the energy of level $n$. Introducing the characteristic vibrational energy defined by ${\mathcal E}_1 \equiv k_{\rm B}\theta_{\rm v}$, and dividing the expression for an upper state $N_{n+m}$ by that of a lower state $N_n$ (where $\Delta {\mathcal E}_{n,m} = {\mathcal E}_{n+m} - {\mathcal E}_n$), yields the Treanor population ratio:
\begin{equation}
    \left.\frac{N_{n+m}}{N_n} \right|_{\rm Tr} = \exp\left( - \frac{m \theta_{\rm v}}{T_{\rm v}} + \frac{m \theta_{\rm v} - \Delta {\mathcal E}_{n,m}/k_{\rm B}}{T_{\rm g}} \right)
    \label{eq:pop_ratio_treanor}
\end{equation}
Substituting the energy gap definition from Eq.~(\ref{eq:energy_gap_anh}) into Eq.~(\ref{eq:pop_ratio_treanor}), the population ratio simplifies compactly using the defect function $\delta(n,m)$:
\begin{equation}
    \left. \frac{N_{n+m}}{N_n} \right|_{\rm Tr} = \exp\left( - \frac{m \theta_{\rm v}}{T_{\rm v}} + \frac{m \theta_{\rm v} \delta(n,m)}{T_{\rm g}} \right)
    \label{eq:treanor_explicit}
\end{equation}
The standard Treanor distribution diverges unphysically at high vibrational levels. In realistic discharges, V-T relaxation overtakes V-V up-pumping, forming a ``Treanor-Gordiets'' distribution\cite{misc:1980:gordiets}. We introduce a flux-corrected Gordiets ratio, ${\cal G}(n,m)$, to analytically model the plateau beyond the Treanor minimum bottleneck, $n^\star$. This function is constructed to enforce a zero population derivative at $n^\star$---where V-V and V-T fluxes balance---and to recover the expected $1/n$ asymptotic decay of high-lying populations, ensuring convergence of the macroscopic energy summation:
\begin{equation}
\left. \frac{N_{n+m}}{N_n} \right|_{\rm Pl} \equiv \mathcal{G}(n,m) \approx \frac{n + \frac{(n^\star)^2}{n}}{n+m + \frac{(n^\star)^2}{n+m}}
\label{eq:plateau_explicit}
\end{equation}
Importantly, the generalized heating formulation derived here is independent of the specific analytic representation of the Gordiets ratio $G(n,m)$. Alternative plateau models may therefore be substituted without modifying the thermodynamic structure of the closure, enabling molecule-specific or higher-fidelity formulations to be incorporated when available. In the present formulation, populations follow the Treanor curve until the Treanor minimum $n^\star$, after which they transition to the flux-corrected plateau. To smoothly capture transitions spanning this minimum without introducing discontinuous piecewise functions, we introduce an effective quantum jump, $m^\star$, defined as:
\begin{equation}
    m^\star \equiv \max\left(0,~ \min\left(m,~n^\star - n\right)\right)
\end{equation}
The Treanor-governed portion of the transition, $m^\star$, becomes $m$ for transitions entirely below the bottleneck, $n^\star - n$ for those crossing it, and zero if the initial state $n$ is already on the plateau. By dynamically truncating the Treanor regime at $n^\star$, the unified population ratio simply becomes the product of the Treanor ratio for the initial jump $m^\star$ and the Gordiets ratio for the remaining jump $m - m^\star$:
\begin{equation}
    \frac{N_{n+m}}{N_n} = \mathcal{G}(n+m^\star, m-m^\star) \exp\left( - \frac{m^\star \theta_{\rm v}}{T_{\rm v}} + \frac{m^\star \theta_{\rm v} \delta(n,m^\star)}{T_{\rm g}} \right) 
    \label{eq:pop_unified}
\end{equation}
To incorporate this kinetic behavior into the energy transfer model, we relate the macroscopic heating and cooling fluxes to the local population ratio. Dividing the definition of the superelastic heating flux (Eq.~\ref{eq:heating_def_anh}) by the cooling flux (Eq.~\ref{eq:level_cooling_def}) and applying the principle of detailed balance (Eq.~\ref{eq:detailed_balance_anh}) yields a direct proportionality:
\begin{equation}
\frac{Q_{\rm v-e}^{(n,m)}}{Q_{\rm e-v}^{(n,m)}} = \frac{N_{n+m}}{N_n} \exp\left(\frac{\Delta {\cal E}_{n,m}}{k_{\rm B} T_{\rm e}}\right)
\label{eq:flux_ratio_identity}
\end{equation}
Expanding the anharmonic energy gap $\Delta {\cal E}_{n,m} = m k_{\rm B} \theta_{\rm v} [1 - \delta(n,m)]$ within the detailed balance exponential leads to the physical flux ratio:
\begin{align}
    \frac{Q_{\rm v-e}^{(n,m)}}{Q_{\rm e-v}^{(n,m)}} &= \frac{N_{n+m}}{N_n} \exp\left(\frac{m\theta_{\rm v}}{T_{\rm e}} - \frac{m\theta_{\rm v}\delta(n,m)}{T_{\rm e}}\right)
    \label{eq:Qve_substituted}
\end{align}
To express this in the standard thermodynamic form, we first note that in the harmonic limit (assuming a Boltzmann distribution), the ratio between the heating and the cooling corresponds to:\cite{pf:2026:parent}
\begin{equation}
    \left.\frac{Q_{\rm v-e}^{(n,m)}}{Q_{\rm e-v}^{(n,m)}}\right|_{\rm har} = \exp\left(\frac{m\theta_{\rm v}}{T_{\rm e}} - \frac{m\theta_{\rm v}}{T_{\rm v}}\right)
\end{equation}
To account for deviations from this ideal baseline, we introduce the generalized anharmonic correction factor $\Phi_{\rm anh}^{(n,m)}$ to enforce the aforementioned kinetic constraints within the final heating formulation:
\begin{equation}
    \frac{Q_{\rm v-e}^{(n,m)}}{Q_{\rm e-v}^{(n,m)}} =   \Phi_{\rm anh}^{(n,m)} 
 \exp\left(\frac{m\theta_{\rm v}}{T_{\rm e}} - \frac{m\theta_{\rm v}}{T_{\rm v}}\right)      
    \label{eq:Qve_unified}
\end{equation}
Equating the latter with Eq.~(\ref{eq:Qve_substituted}) and isolating the anharmonic correction factor leads to:
\begin{equation}
\Phi_{\rm anh}^{(n,m)} = \frac{N_{n+m}}{N_n} \exp\left( \frac{m\theta_{\rm v}}{T_{\rm v}} - \frac{m\theta_{\rm v} \delta(n,m)}{T_{\rm e}} \right)
\label{eq:phi_anh_intermediate}
\end{equation}
To express this entirely in macroscopic thermodynamic variables without relying on explicit state populations, we substitute the unified Treanor-Gordiets ratio outlined in Eq.~(\ref{eq:pop_unified}) into this relation. Dividing by the harmonic scaling term groups the exponential arguments directly, yielding the final closed-form correction factor:
\begin{align}
\Phi_{\rm anh}^{(n,m)} &= \mathcal{G}(n+m^\star, m-m^\star) \nonumber \alb &\times \exp\left( \theta_{\rm v} \left[ \frac{m - m^\star}{T_{\rm v}} + \frac{m^\star \delta(n,m^\star)}{T_{\rm g}} - \frac{m\delta(n,m)}{T_{\rm e}} \right] \right)
\label{eq:phi_anh}
\end{align}
Isolating the channel-specific heating term $Q_{\rm v-e}^{(n,m)}$ in Eq.~(\ref{eq:Qve_unified}) and summing over all channels $m$ and all vibrational levels $n$, the total heating rate becomes:
\begin{equation}
    Q_{\rm v-e} = \sum_{n=0}^{\infty} \sum_{m=1}^{\infty}  Q_{\rm e-v}^{(n,m)} \Phi_{\rm anh}^{(n,m)}
    \exp\left(\frac{m\theta_{\rm v}}{T_{\rm e}} - \frac{m\theta_{\rm v}}{T_{\rm v}}\right)  
    \label{eq:Qve_final}
\end{equation}
This formulation generalizes our thermodynamically consistent model to Treanor-Gordiets distributions. It should be noted that the assumption of a Maxwellian electron energy distribution (utilized in Eq.~(\ref{eq:detailed_balance_anh}) to analytically ensure compliance with the Second Law of Thermodynamics) will generally fail in plasmas with high electron temperatures or strong electric fields, where the high-energy tail becomes non-Maxwellian. Nevertheless, the generalized framework in Eq.~(\ref{eq:Qve_final}) is expected to remain highly useful in such regimes. For non-Maxwellian plasmas, the Maxwellian superelastic heating term, $Q_{\rm e-v}^{(n,m)} \exp(m\theta_{\rm v}/T_{\rm e} - m\theta_{\rm v}/T_{\rm v})$, can simply be substituted with a non-Maxwellian rate computed via a Boltzmann solver (e.g., LoKI, BOLSIG+). The anharmonic correction factor ($\Phi_{\rm anh}$) is anticipated to remain a robust approximation; its internal dependence on the electron distribution scales with $1/T_{\rm e}$, rendering it negligible at the elevated temperatures where non-Maxwellian deviations are most pronounced. A rigorous verification of this non-Maxwellian decoupling is deferred to subsequent work.

To provide an analytical closure for non-equilibrium electron temperature evolution, the generalized solution presented in Eq.~(\ref{eq:Qve_final}) inherently depends on the critical crossover point, $n^\star$, which is now determined.

Physically, the crossover point $n^\star$ must correspond to the Treanor minimum bottleneck, where the local population derivative vanishes. We analytically locate $n^\star$ by taking the continuous limit ($m \to 0$) of the pure Treanor ratio (Eq.~\ref{eq:treanor_explicit}). Setting the exponential argument to zero yields the transition criterion $\delta(n,0) = T_{\rm g}/T_{\rm v}$, where the continuous anharmonic defect function evaluated at $m=0$ is given by:
\begin{equation}
    \delta(n,0) = \frac{x_{\rm e}(2n - 1) - y_{\rm e}(3n^2 + 3n - 2.5)}{1 - 2x_{\rm e} + 3.25y_{\rm e}}
\end{equation}
Equating this to the temperature ratio results in a quadratic equation for the bottleneck level. Solving for the vibrational level and selecting the negative root to preserve physical bounds yields the critical crossover point, $n^\star$:
\begin{equation}
    n^\star = \frac{2x_{\rm e} - 3y_{\rm e} - \sqrt{4x_{\rm e}^2 - 12y_{\rm e} \left[ \frac{T_{\rm g}}{T_{\rm v}} + (2x_{\rm e} - 3.25y_{\rm e})\left(1 - \frac{T_{\rm g}}{T_{\rm v}}\right) \right]}}{6y_{\rm e}}
    \label{eq:n_star_def}
\end{equation}
Taking the mathematical limit of Eq.~(\ref{eq:n_star_def}) as $y_{\rm e} \to 0$ yields a corrected linear minimum, $n^\star = T_{\rm g}/(2x_{\rm e} T_{\rm v}) - T_{\rm g}/T_{\rm v} + 1/2$. Because $x_{\rm e}$ is small, this naturally reduces to the classical Treanor formulation, $n^\star \approx T_{\rm g}/(2x_{\rm e} T_{\rm v}) + 1/2$, for most practical computations. However, by retaining higher-order $y_{\rm e}$ terms, Eq.~(\ref{eq:n_star_def}) locates the bottleneck deep into the manifold without unphysical kinetic shifts. To prevent a negative square root, Eq.~(\ref{eq:n_star_def}) is applied only when $T_{\rm g} < T_{\rm v}$; otherwise, $n^\star$ is set safely beyond the manifold's truncation limit.

To further verify physical accuracy, we demonstrate that the formulation satisfies strict thermodynamic consistency at complete thermal equilibrium:
\begin{equation}
    T_{\rm e} = T_{\rm v} = T_{\rm g} = T
\end{equation}
Substituting these equalities into the total heating rate expression in Eq.~(\ref{eq:Qve_final}), we examine the two temperature-dependent components individually. First, the harmonic scaling term vanishes because $T_{\rm e} = T_{\rm v}=T$:
\begin{equation}
    \left.\exp\left(\frac{m\theta_{\rm v}}{T_{\rm e}} - \frac{m\theta_{\rm v}}{T_{\rm v}}\right)\right|_{T} = \exp(0) = 1
\end{equation}
Second, we examine the anharmonic correction factor $\Phi_{\rm anh}^{(n,m)}$ from Eq.~(\ref{eq:phi_anh}). Under strict thermal equilibrium ($T_{\rm v} = T_{\rm g} = T$), the Treanor minimum $n^\star$ is pushed beyond the dissociation limit. Consequently, all physically bound transitions occur well below the bottleneck ($n+m \ll n^\star$).

Because the transition manifold is confined entirely to the pure Treanor regime, the effective quantum jump simplifies to $m^\star = m$. Substituting this alongside the thermal equilibrium temperatures into Eq.~(\ref{eq:phi_anh}) yields:
\begin{equation}
\left.\Phi_{\rm anh}^{(n,m)}\right|_{T} = \mathcal{G}(n+m, 0) \exp\left( \frac{\theta_{\rm v}}{T} \big[ 0 + m \delta(n,m) - m\delta(n,m) \big] \right)
\end{equation}
Since the Gordiets ratio for a zero-quantum jump is trivially unity ($\mathcal{G}(n+m, 0) = 1$) and the anharmonic defect terms in the exponent perfectly cancel, this reduces to:
\begin{equation}
\left.\Phi_{\rm anh}^{(n,m)}\right|_{T} = 1 \cdot \exp(0) = 1
\end{equation}
\begin{figure*}[!h]
     \centering
     \subfigure[]{\includegraphics[width=0.48\textwidth]{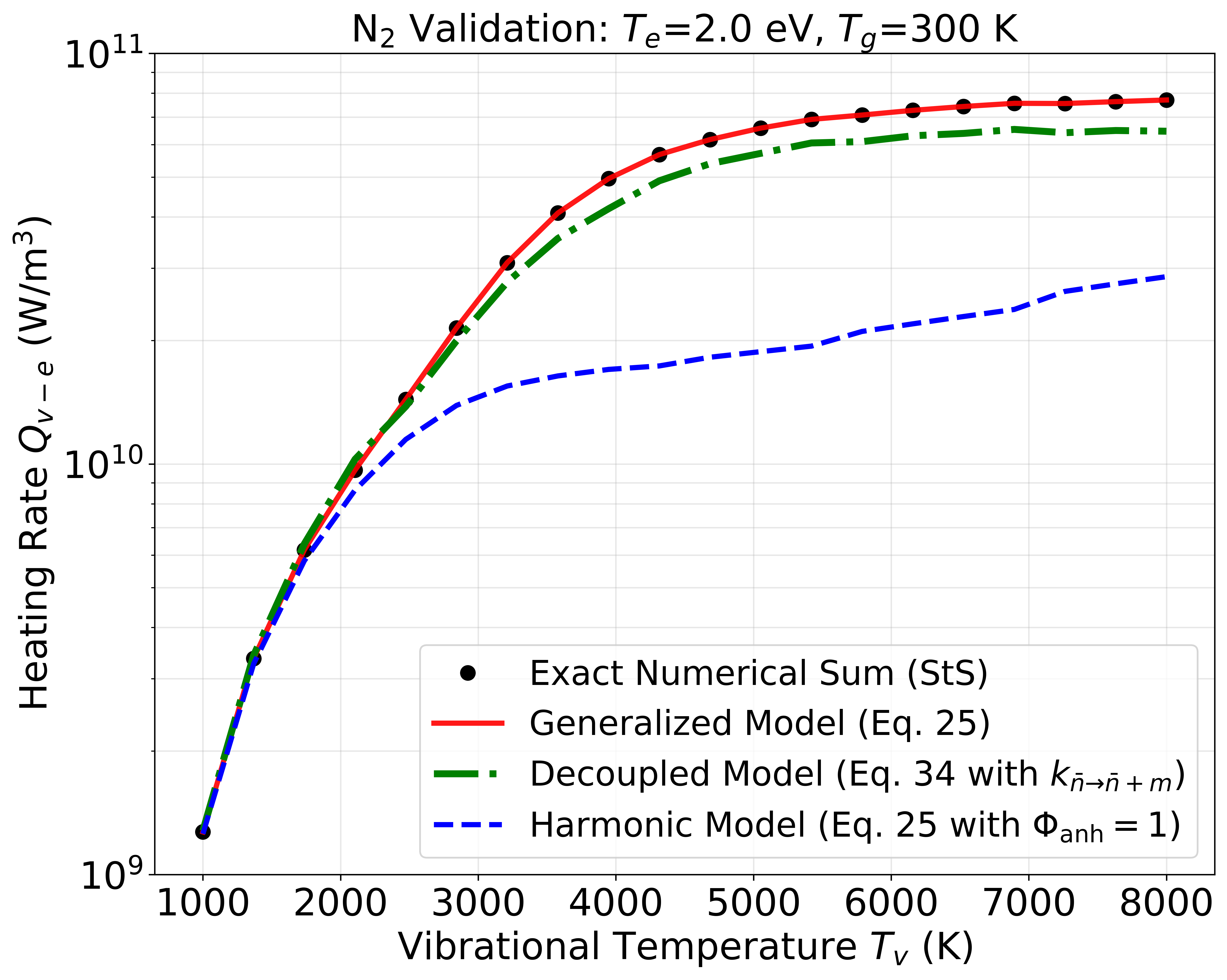}}
     \subfigure[]{\includegraphics[width=0.48\textwidth]{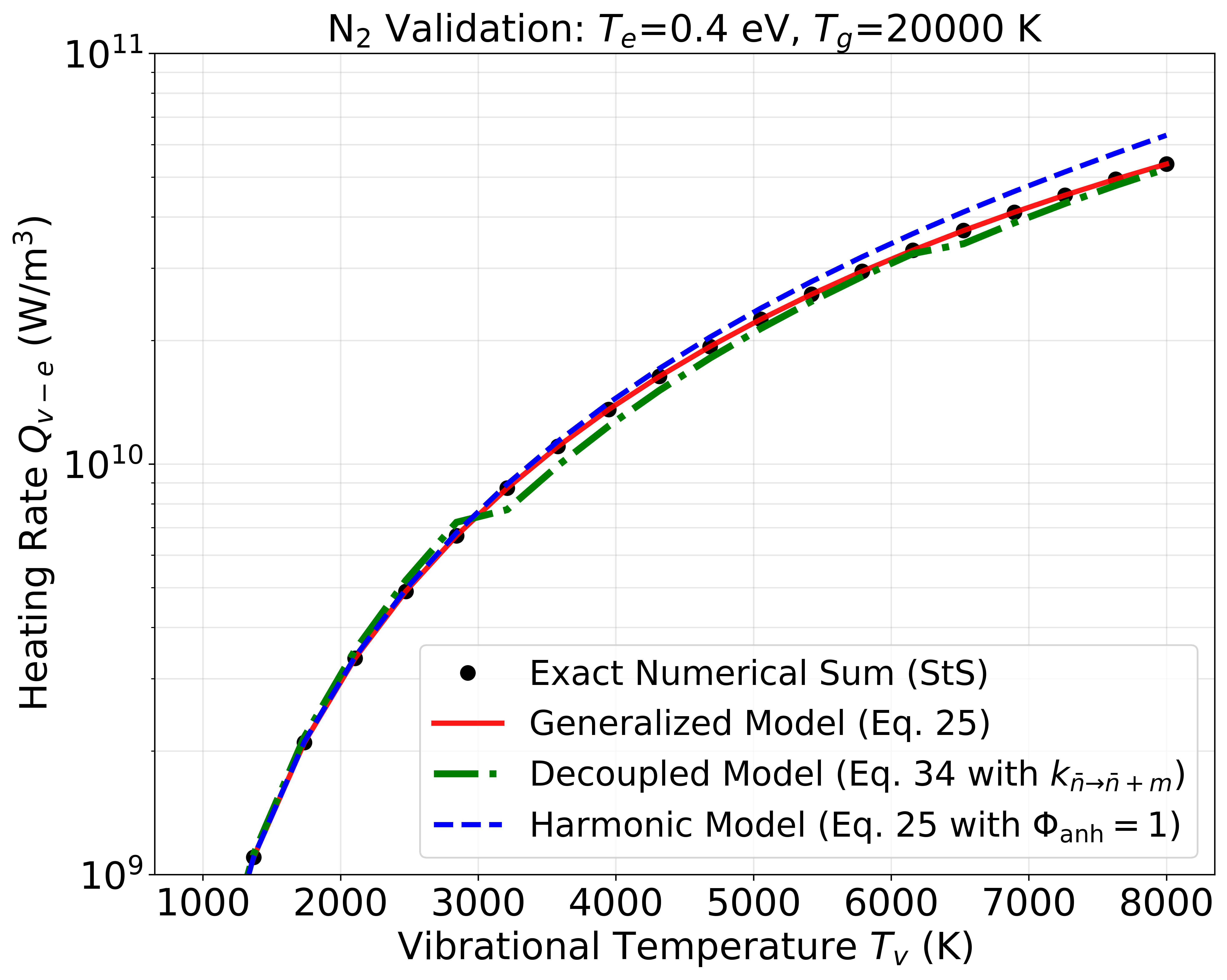}}
     \subfigure[]{\includegraphics[width=0.48\textwidth]{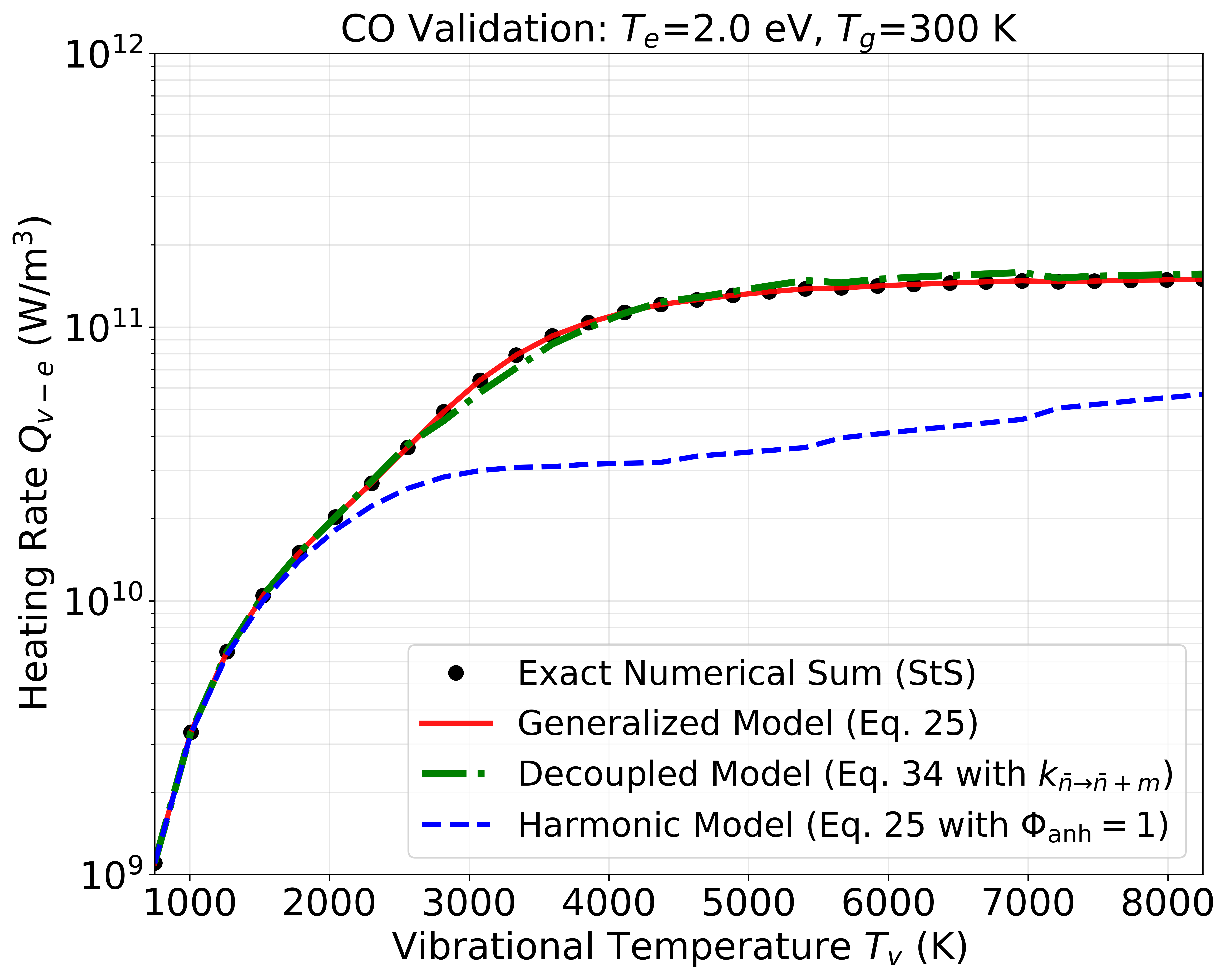}}
     \subfigure[]{\includegraphics[width=0.48\textwidth]{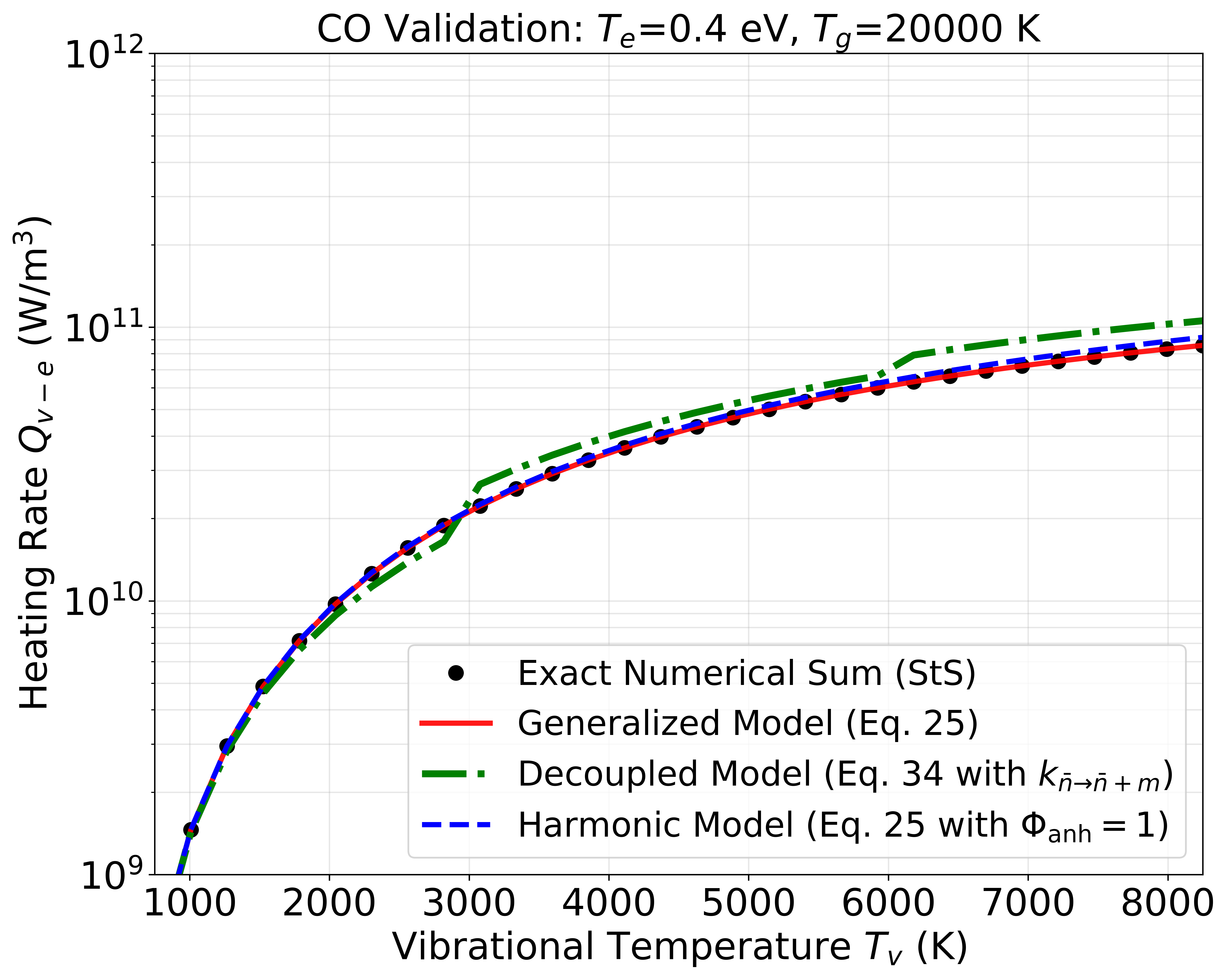}}  
     \figurecaption{Comparison of vibrational-electron heating rates for (a) N$_2$ at $T_{\rm e}=2.0$~eV and $T_{\rm g}=300$~K, (b) N$_2$ at $T_{\rm e}=0.4$~eV and $T_{\rm g}=20000$~K, (c) CO at $T_{\rm e}=2.0$~eV and $T_{\rm g}=300$~K, and (d) CO at $T_{\rm e}=0.4$~eV and $T_{\rm g}=20000$~K.}
     \label{fig:validation}
\end{figure*}
Substituting these results back into Eq.~(\ref{eq:Qve_final}), the expression for the total heating rate simplifies to the sum of the level-specific cooling fluxes:
\begin{equation}
    \left.Q_{\rm v-e}\right|_{T} = \sum_{m=1}^\infty \sum_{n=0}^{\infty} \left[ Q_{\rm e-v}^{(n,m)} \cdot 1 \cdot 1 \right] = \sum_{m=1}^\infty \sum_{n=0}^{\infty} Q_{\rm e-v}^{(n,m)}
\end{equation}
By definition, the term $Q_{\rm e-v}^{(n,m)}$ is the cooling flux for a specific transition (Eq.~\ref{eq:level_cooling_def}). Summing these fluxes over all levels $n$ yields the macroscopic cooling rate per channel $Q_{\rm e-v}^{(m)}$ (Eq.~\ref{eq:cooling_def_anh}), and summing over all channels $m$ recovers the total cooling rate $Q_{\rm e-v}$ (Eq.~\ref{eq:total_cooling_anh}). Thus, we recover the strict equality:
\begin{equation}
    Q_{\rm v-e} = Q_{\rm e-v}\quad \textrm{when} \quad T_{\rm v}=T_{\rm e}=T_{\rm g}
\end{equation}
This proves that the generalized model is thermodynamically consistent. It ensures zero net energy transfer when the electrons are  thermalized with both the vibrational and translational modes of the gas.

To efficiently implement the exact state-resolved closure (Eq.~\ref{eq:Qve_final}) in macroscopic fluid models, we derive a decoupled form by assuming the forward electron-impact rate coefficient is independent of the initial level ($k_{n \rightarrow n+m} \approx k_m$). Using the anharmonic energy gap (Eq.~\ref{eq:energy_gap_anh}), the level-specific cooling flux expands as $Q_{\rm e-v}^{(n,m)} \approx N_{\rm e} N_n k_m m k_{\rm B} \theta_{\rm v} [1 - \delta(n,m)]$. By introducing the fractional population $X_n = N_n / N$ (with $N$ the total molecular number density) and a harmonic-equivalent macroscopic cooling rate $Q_{\rm e-v}^{m}|_{\rm har} = N_{\rm e} N m k_{\rm B} \theta_{\rm v} k_m$, we factor the $n$-independent terms out of Eq.~(\ref{eq:Qve_final}):
\begin{equation}
 Q_{\rm v-e} \approx \sum_{m=1}^{\infty} Q_{\rm e-v}^{m}|_{\rm har} \exp\left( \frac{m\theta_{\rm v}}{T_{\rm e}} - \frac{m\theta_{\rm v}}{T_{\rm v}} \right) \sum_{n=0}^{\infty} X_n [1 - \delta(n,m)] \Phi_{\rm anh}^{(n,m)}
\label{eq:Qve_decoupled}
\end{equation}
This decoupled formulation cleanly separates electron collision cross-sections from the thermodynamic state of the vibrational manifold. The outer summation drives the energy transfer for channel $m$, while the inner summation isolates the anharmonic physics into a dimensionless thermodynamic weighting factor. By reducing the kinetic burden from $m \times n$ reaction rates to just $m$ rates, this approximation yields a 40- to 70-fold reduction in rate evaluations, significantly enhancing computational efficiency.

We now proceed to validate our generalized governing equation against a direct state-to-state (StS) summation. This requires determining the population densities across the entire vibrational manifold. By evaluating the generalized population ratio (Eq.~\ref{eq:pop_unified}) relative to the ground state---setting the initial state to zero and the quantum jump to $n$---we bypass the need for piecewise definitions for all excited states ($n > 0$). The population density of the $n$th vibrationally excited state then condenses into a single, continuous analytical expression governed by the effective Treanor level, $n_{\cal T} = \min(n, n^\star)$:
\begin{equation}
    N_n = N_0 \, \mathcal{G}(n_{\cal T}, n-n_{\cal T}) \exp\left( -\frac{n_{\cal T} \theta_{\rm v}}{T_{\rm v}} + \frac{n_{\cal T} \theta_{\rm v} \delta(0,n_{\cal T})}{T_{\rm g}} \right)
    \label{eq:Nn_unified}
\end{equation}
where $N_0$ is the population density of the ground state.

We evaluate representative $\rm N_2$ and CO plasmas across two distinct regimes: a cold-gas case ($T_{\rm e} = 2.0$ eV, $T_{\rm g} = 300$ K; Fig.~\ref{fig:validation}a,c) and a hot-gas case ($T_{\rm e} = 0.4$ eV, $T_{\rm g} = 20000$ K; Fig.~\ref{fig:validation}b,d). The former simulates the strong Treanor pumping characteristic of the inter-pulse stages of plasma-assisted combustion, while the latter represents the thermodynamic depopulation driven by strong shock heating in hypersonic flows. In both, $T_{\rm v}$ spans highly non-equilibrium states. The population $N_n$ is constructed using the Treanor-Gordiets formulation (Eq.~\ref{eq:Nn_unified}) with $n_{\max}=45$ for $\rm N_2$ and $80$ for CO where $n_{\max}$ is the truncation limit of the vibrational manifold. We assume realistic discharge conditions: $N_{\rm e} = 10^{19}$~m$^{-3}$ and $N \approx 10^{25}$~m$^{-3}$ (approx. 0.4 atm). Forward excitation rates $k_{n \to n+m}$ are derived from state-to-state cross-sections by Laporta \textit{et al.} for $\rm N_2$~\cite{psst:2014:laporta} and CO.\cite{psst:2016:laporta}

We compute the superelastic heating rate $Q_{\rm v-e}$ using four methods: (1) a brute-force StS benchmark summing detailed-balance contributions across all specific channels ($n+m \to n$) with actual populations and anharmonic gaps; (2) an harmonic model (Eq.~\ref{eq:Qve_final} with $\Phi_{\rm anh}=1$); (3) our generalized closed-form model incorporating the anharmonic correction $\Phi_{\rm anh}$ (Eq.~\ref{eq:Qve_final}); and (4) the decoupled macroscopic model (Eq.~\ref{eq:Qve_decoupled}). For the latter, we evaluate the channel-specific harmonic cooling rate $Q_{\rm e-v}^m|_{\rm har}$ efficiently by calculating the forward rate at a dynamically weighted representative level, $\bar{n} = \sum_{n=0}^{n_{\max}} n X_n$ with $X_n$ the molar fraction of the $n$th vibrational population. Setting $k_m \approx k_{\bar{n} \to \bar{n}+m}$ (where $k_{\bar{n} \to \bar{n}+m}$ is the forward rate evaluated at this representative level) analytically captures the kinetic shift of the reacting population without requiring a full state-dependent cross-section matrix.

Figure~\ref{fig:validation} highlights both temperature regimes. At low gas temperatures (Fig.~\ref{fig:validation}a,c) and $T_{\rm v} \approx T_{\rm g}$, distributions are Boltzmann-like, and all four models converge. As $T_{\rm v}$ exceeds 2000~K, Treanor effects drive substantial high-energy overpopulation.  While the harmonic model evaluates cooling fluxes using the correct non-equilibrium populations, enforcing $\Phi_{\rm anh}=1$ imposes equal energy spacing between vibrational levels. This harmonic assumption underestimates the detailed-balance scaling for reverse superelastic heating, thereby undercounting de-excitations from overpopulated, high-lying states.  Consequently, it underpredicts the heating rate by approximately a factor of five compared to the StS exact solution (dashed blue line). In this $T_{\rm v} > T_{\rm g}$ regime, our anharmonic correction $\Phi_{\rm anh} > 1$ naturally accounts for the superelastic surplus. The dynamically weighted decoupled model (green dash-dotted line) closely matches the exact summation; by tracking population migration toward the high-energy plateau, it scales collision efficiency accurately and avoids overpredictions inherent to ground-state-anchored rates.

Conversely, at high gas temperatures ($T_{\rm g} > T_{\rm v}$; Fig.~\ref{fig:validation}b,d), the Treanor-Gordiets population ratio $N_{n+m}/N_n$ drops below the standard $\exp(-m\theta_{\rm v}/T_{\rm v})$ macroscopic scaling. While the harmonic baseline uses correct non-equilibrium densities $N_n$ for forward cooling fluxes, enforcing $\Phi_{\rm anh}=1$ mathematically overestimates the superelastic return flux. This causes the harmonic model to artificially overpredict total heating by approximately 15\%. Our generalized model captures this physics perfectly: $\Phi_{\rm anh}$ drops below unity to properly scale down heating fluxes. The decoupled model also maintains high fidelity; as the upper plateau depopulates, the dynamic representative level $\bar{n}$ shifts back toward fundamental levels, ensuring macroscopic rates remain physically representative.

While the generalized model captures anharmonic physics via $\Phi_{\rm anh}$ to reproduce the StS summation with negligible error, the true practical advantage lies in its decoupled macroscopic counterpart. Combining this analytic formulation with dynamic representative transition rates achieves a 40- to 70-fold reduction in computational effort related to rate evaluations. It maintains high fidelity---deviating by 15\% at most---while circumventing the severe underpredictions of the harmonic baseline. Furthermore, depending on far fewer reactions makes this decoupled model highly advantageous compared to a full state-to-state approach, which rarely possesses the complete set of required cross-sections across all temperature ranges. Ultimately, pairing an anharmonic closure with representative rates provides a robust and computationally efficient path toward reduced-order plasma models for macroscopic fluid solvers.

\section*{Data Availability}

The data that support the findings of this study are available from the author upon reasonable request.

\section*{Acknowledgements}

The author extends heartfelt acknowledgement to the reviewers for their inspiring guidance. Although several of the comments were graciously presented as minor and optional, the perspective on the physical limitations of current StS frameworks struck a chord. This shared philosophy provided the crucial impetus to push the paper significantly beyond its original scope. In addition, the author wishes to acknowledge Vincenzo Laporta for providing the rate data for CO e-V cooling.

%\appendix
%\section{Appendix Test}
\footnotesize
\bibliography{all}
\bibliographystyle{aiaa2}
\end{document}